\documentclass[3p,times]{elsarticle}

\usepackage{ecrc}

\volume{00}

\firstpage{1}

\journalname{Nuclear Physics A}

\runauth{}

\jid{procs}

\jnltitlelogo{Nuclear Physics A}

\usepackage{amssymb}

\usepackage[figuresright]{rotating}

\begin{document}

\begin{frontmatter}



\dochead{}

\title{First measurements of beauty quark production \\at $\sqrt{s}=7$ TeV with the CMS experiment}


\author{Vincenzo Chiochia\\{\it on behalf of the CMS Collaboration}}

\address{Universit\"at Z\"urich, Physik-Institut, Winterthurerstr. 190, 8057 Z\"urich, Switzerland}

\begin{abstract}
This article summarizes the first measurements of inclusive beauty production cross section in proton-proton collisions at  $\sqrt{s}=7$~TeV and central rapidities. The results are based on different techniques, such as the identification of semileptonic $b$-decays into muons and inclusive jet measurements with secondary vertex tagging. The measurements probe $b$-quark production in different  regions of transverse momenta. The experimental results are compared with next-to-leading order QCD predictions and various Monte Carlo models.
\end{abstract}


\end{frontmatter}


\section{Introduction\label{sec:Intro}}
It is important to understand inclusive $b$-quark production at LHC experiments for various reasons. Firstly, in order to test QCD predictions which have been computed at next-to-leading order (NLO) precision but are still characterized by large scale dependence. Secondly, because $b$-jets represent an important source of background for many of the most interesting physics searches, as the Higgs boson and Supersymmetric extensions of the Standard Model. Most recent measurements performed at Tevatron~\cite{Abachi:1994kj,Abbott:1999se,Abe:1993sj,Aaltonen:2009xn}, HERA~\cite{Adloff:1999nr,Aktas:2005zc,Chekanov:2008tx,Chekanov:2009kj} and LEP~\cite{Acciarri:2000kd} are in reasonable agreement with QCD predictions in most regions of the phase space. 

Two preliminary measurements of inclusive $b$-quark production at the center-of-mass energy of 7~TeV have been performed with the CMS experiment~\cite{:2008zzk}, based on different experimental techniques. The results are obtained with data collected in March-July 2010. A first measurement is based on the identification of semileptonic decays of $b$ quarks into muons and jets. Muons from $b$- and $c$-quark decays can be distinguished using the transverse momentum relative to the jet, which is on average larger for $b$-events than in $c$-decays and for muons from light hadrons. This measurement probes the production process at low transverse momenta. A second measurement is based on the reconstruction of the secondary vertex in jets from the $B$-hadron decays, exploiting the high spatial resolution of the silicon pixel tracker. This measurement extends to large $b$-quark transverse momenta. Both results are compared to NLO QCD predictions and various Monte Carlo (MC) models.

\section{Measurement techniques and results\label{sec:Results}}


\subsection{Cross section measurement with semi-leptonic decays}

The measurement is based on the reconstruction of the muon from the semi-leptonic $b$-decay and associated jet~\cite{BPH-10-007}, using an integrated luminosity $L=8.1$ nb$^{-1}$. At least one well-reconstructed muon with transverse momentum $p_{T,\mu}> 6$ GeV and pseudorapidity $|\eta_\mu|< 2.1$ is required. Further cuts are applied on the longitudinal impact parameter, on the minimal number of hits associated to the track on on the quality of the track fit. Tracks with $p_T>300$ MeV are clustered into {\it track-jets} with the anti-$k_T$ jet algorithm and $R=0.5$. The jet is defined as {\it b-jet} if it contains a muon satisfying the above requirements. After subtracting the muon momentum from the track-jet momentum, the track-jet energy is required to be $E_T> 1$ GeV in the plane transverse to the beam line.

From the momenta of the selected muon ($\vec{p_\mu}$) and the associated track jet ($\vec{p_j}$), the relative transverse momentum of the muon with respect to its track jet is calculated as $p^{rel}_T = |\vec{p_\mu}\times \vec{p_j}|/|\vec{p_\mu}|$. A fit to the observed $p^{rel}_T$ spectrum, based on templates obtained from simulation (signal and part of the background) and data (the remaining background), is used to determine the fraction of signal events among all events passing the event selection. The templates used in the fitting algorithm are determined separately for the full sample and for each bin in muon transverse momentum and pseudorapidity. Since the shape of the $p^{rel}_T$ distribution from charm decays and hadrons from light quarks or gluons cannot be distinguished by the fit, the two background components are combined.

\begin{figure}[htb!]
	\begin{center}
	\includegraphics[width=10cm]{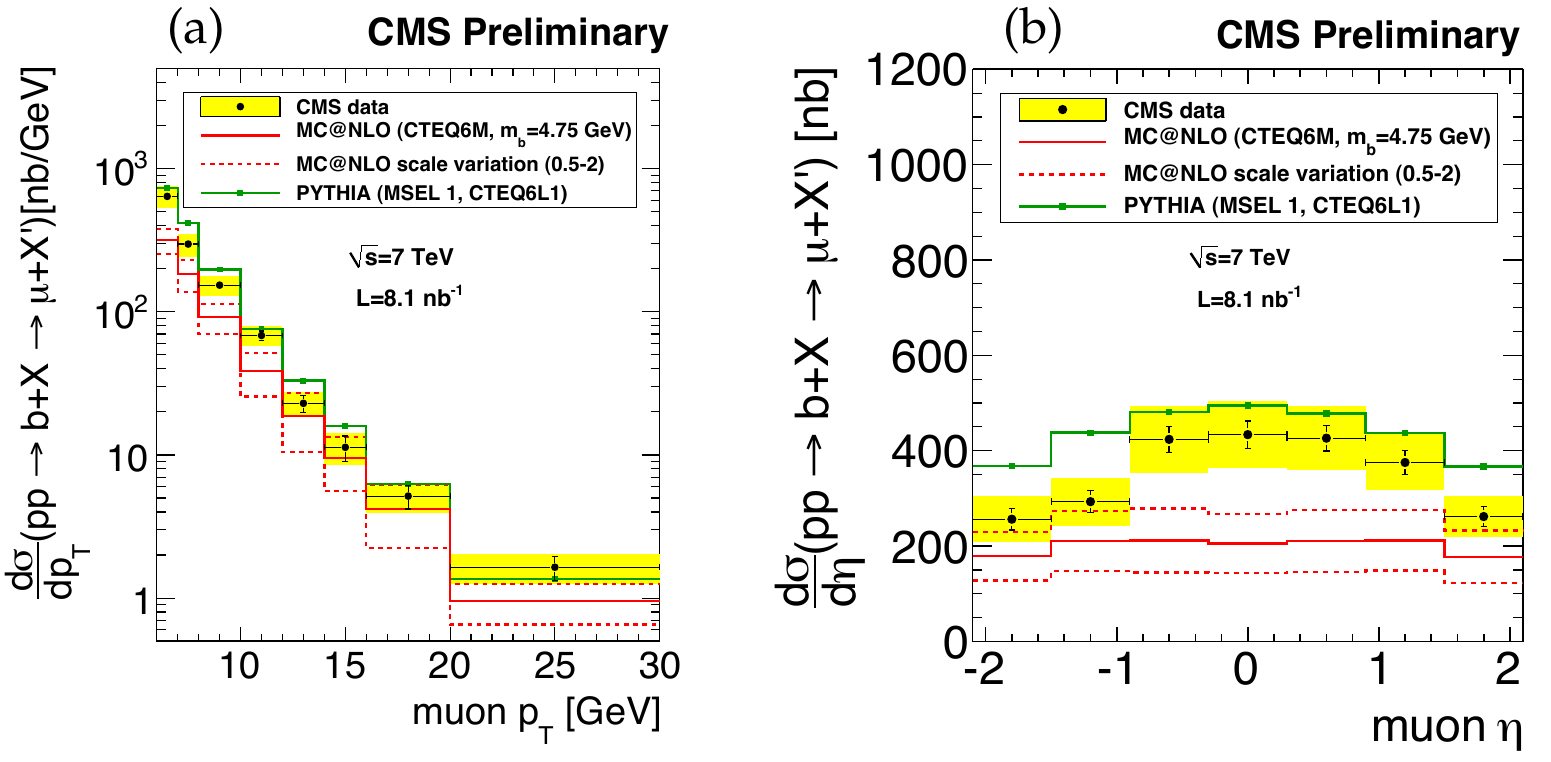}
	\vspace{-0.5cm}
	\includegraphics[width=6cm]{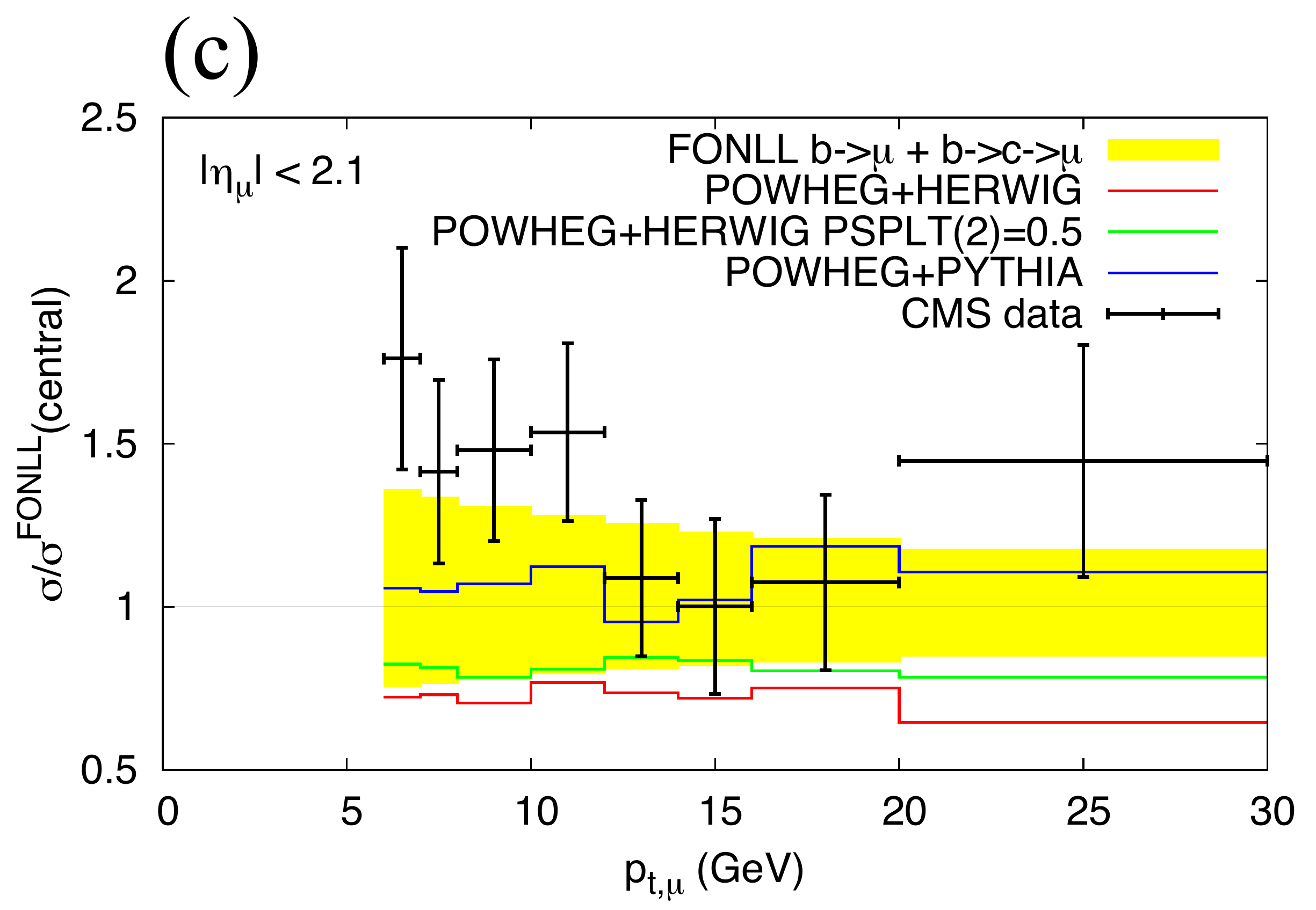}
	\end{center}
	\caption{Differential $b$ cross section as function of the muon transverse momentum for $|\eta_\mu|< 2.1$ (a) and pseudorapidity for $p_{T,\mu}> 6$ GeV (b), compared with PYTHIA and MC@NLO predictions. The yellow band shows the quadratic sum of statistical and systematic errors (the uncertainty on the luminosity measurement is not included). The ratio of the measured cross section to the theoretical expectations from FONLL and POWHEG are shown in (c).\label{fig:ptreldiffxsec}}
\end{figure}

The inclusive $b$-quark production cross section, $\sigma_b$, is calculated from $\sigma_b = N_b/(\epsilon L)$, where $N_b$ is the number of events from $b$-decays extracted from the fit, $\epsilon$ is the overall event selection efficiency and $L$ the integrated luminosity. The result of the inclusive b-quark production cross section to muons, for the visible range $p_{T,\mu}>6$ GeV and $|\eta_\mu|<2.1$ is $\sigma_b=(1.48 \pm 0.04_{\rm stat} \pm 0.22_{\rm syst} \pm 0.16_{\rm lumi})\; \mu$b. Single differential cross sections as function of the muon transverse momentum and pseudorapidity are obtained by determining $N_b$ and the efficiency in each bin (see Fig.~\ref{fig:ptreldiffxsec}). The systematic uncertainties (16\%-20\%) are dominated by the description of the background from light quarks and gluons and modeling of the underlying event. At the present early stage of the CMS experiment, the integrated luminosity recorded is known to about 11\% precision.

Theoretical predictions for the cross section measurement were obtained with PYTHIA 6.4~\cite{Sjostrand:2006za}, HERWIG 6.5~\cite{Corcella:2002jc}, MC@NLO 3.4 ~\cite{Frixione:2003ei}, FONLL \cite{Cacciari:2003uh} and POWHEG~\cite{Frixione:2007vw}. The CTEQ6L1 and CTEQ6M parton densities~\cite{Pumplin:2002vw} were used for PYTHIA and MC@NLO predictions, respectively. The PYTHIA prediction for the visible $b$-quark cross section is $\sigma_{\rm PYTHIA}=1.8~\mu$b, while MC@NLO gives $[0.84^{+0.36}_{-0.19}({\rm scale})\pm0.08(m_b)\pm0.04({\rm PDF})]~\mu$b. The error for MC@NLO is obtained by changing the QCD renormalization and factorization scales independently from half to twice their default values. The PYTHIA and MC@NLO predictions for the differential cross sections are shown in Fig.~\ref{fig:ptreldiffxsec}(a)-(b). While PYTHIA predictions are generally in agreement with the measurements, MC@NLO is below the measurement at low transverse momenta and central pseudorapidities. The HERWIG calculation with massive quarks agrees with the MC@NLO prediction within the theoretical uncertainties. The ratio of the measured differential cross section as function of $p_{T,\mu}$ divided by the FONLL and POWHEG predictions are shown in Fig.~\ref{fig:ptreldiffxsec}(c). The POWHEG matrix element calculation is interfaced both to the PYTHIA and HERWIG parton shower. The FONLL calculation generally agrees with the data with the larger difference in the lowermost $p_{T,\mu}$ bin. The POWHEG calculation with PYTHIA parton shower is in agreement with FONLL. The POWHEG prediction is below the data if interfaced with the HERWIG parton shower.


\subsection{B-jet cross section measurement}

An additional measurement is performed, based on finding the decay vertex of $B$ hadrons within jets~\cite{BPH-10-009}. Secondary vertices with at least three associated tracks and hits in the silicon pixel detector provide a clean signal against backgrounds from light quark and gluon jets. The secondary vertices from $b$- and $c$-quark decays can be distinguished by their relative distance from the primary vertex using a 3D decay length significance, which is higher for $b$-jets than for $c$- and light flavor jets.

The inclusive jet data is collected using a combination of minimum bias and single jet triggers. The jets with transverse momentum in the range $18<p_T<300$~GeV and rapidity $|y|<2$ are reconstructed with the anti-kT algorithm~\cite{Cacciari:2008gp}, with the jet clustering using a distance parameter $R=0.5$. Particle Flow objects~\cite{PFT-09-001,PFT-10-001} are utilized as input to the clustering algorithm, allowing for a reliable jet energy reconstruction and good energy resolution down to low transverse jet momenta. Jets from $b$-decays are identified using a secondary vertex high-purity tagger~\cite{BTV-10-001}. The secondary vertex is fitted with at least three charged particle tracks. A selection on the reconstructed 3D decay length significance is applied, corresponding to about 0.1\% efficiency to tag light flavor jets and 60\% efficiency to identify $b$ jets at $p_T=100$~GeV. The $b$ identification efficiency with the selections used in theis analysis is between 6\% and 60\% at $p_T>18$~GeV and $|y|<2$. The efficiency rises at higher $p_T$ as the $b$-hadron decay time increases in the laboratory frame, facilitating the identification of the decay vertex.

The production cross section for $b$ jets is calculated as a double differential $d\sigma/(dp_Tdy)=N_t f_b C/(\epsilon_{\rm jet} \epsilon_b \Delta p_T \Delta y L)$, where $N_t$ is the measured number of tagged jets per bin, $\Delta p_T$ and $\Delta y$ are the bin widths in $p_T$ and $y$, $f_b$ is the fraction of tagged jets containing a $B$ hadron, $\epsilon_b$ is the $b$ tagging efficiency, $\epsilon_{\rm jet}$ is the jet reconstruction efficiency and $C$ is the unfolding correction. The integrated luminosity, $L$, is 60~nb$^{-1}$. The $\epsilon_{\rm jet}$,  $\epsilon_b$ and $f_b$ are all calculated from MC in bins of reconstructed $p_T$ and $y$. The b-tagged sample purity was also estimated from data, using template fits
to the secondary vertex mass distribution, and the results were found to be in good agreement
with MC expectations, well within the 3\% statistical uncertainty. This constrains the charm
mistag rate to within 20\% of the MC expectation. The correction factor $C$ unfolds the measured $p_T$ back to particle level using the {\it ansatz method}~\cite{QCD-10-011}.

\begin{figure}[htb!]
	\begin{center}
	\includegraphics[width=13cm]{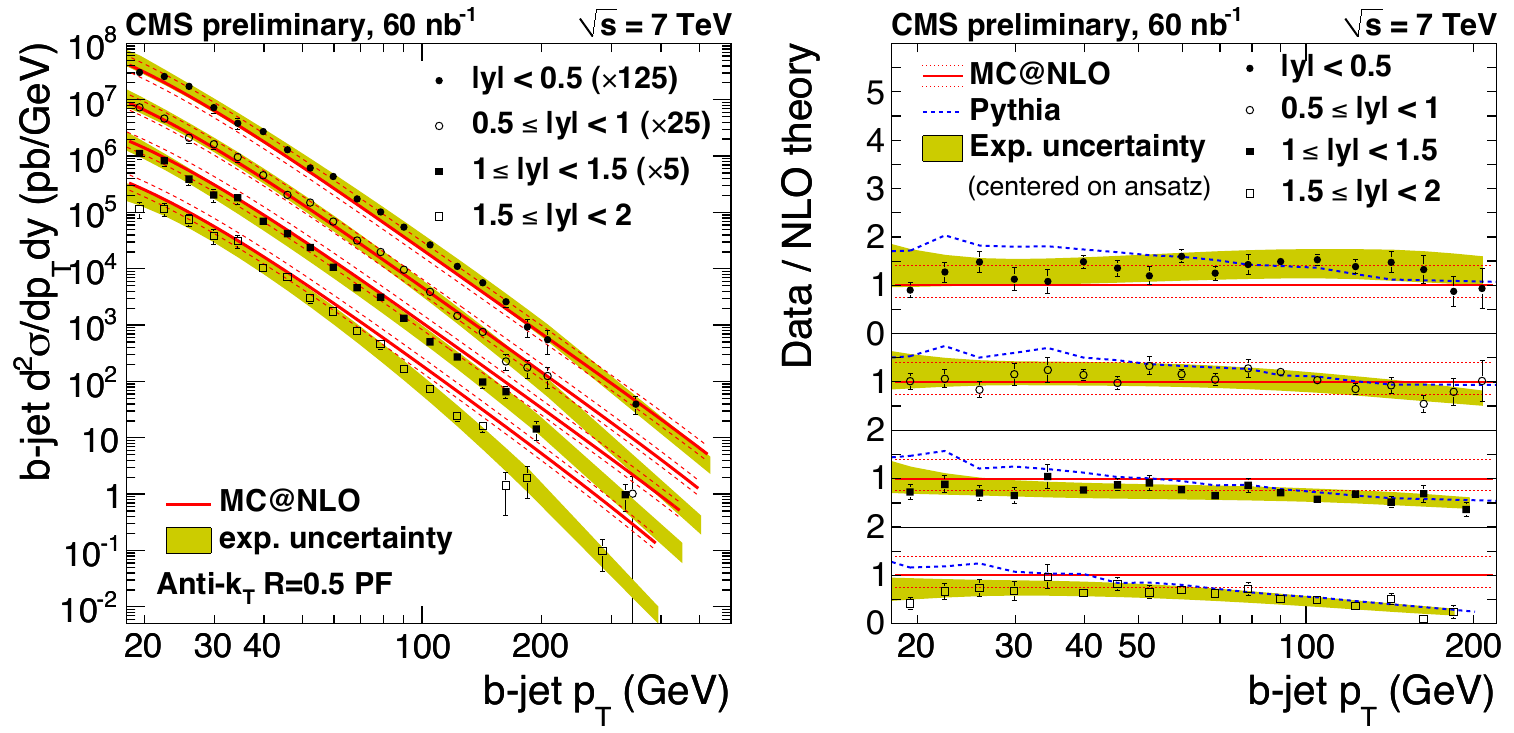}
	\vspace{-0.5cm}
	\end{center}
	\caption{Left: Differential $b$-jet cross section as function of the transverse momentum in different rapidity regions. Right: Ratio of the measured cross section to the NLO QCD prediction.\label{fig:bjetdiffxsec}}
\end{figure}

The measured $b$-jet cross section is shown in Fig.~\ref{fig:bjetdiffxsec} as function of the jet $p_T$, in different rapidity bins. The leading systematic uncertainties at $p_T>30$~GeV are from the $b$-jet energy scale relative to inclusive jets (4Ð5\%), from the data-driven constraints on b-tagging efficiency (20\%) and from the mistag rate uncertainty for charm jets (3Ð4\%) and for light flavor jets (1-10\%).

The inclusive $b$-jet prediction is calculated with MC@NLO, using the CTEQ6M PDF set and the $b$-quark mass of 4.75 GeV. The results are compared to a NLO theory prediction (MC@NLO) and to the Pythia MC, and are found to be in good agreement with Pythia and in reasonable agreement with MC@NLO. The total $b$-quark cross section of 238~$\mu$b from the MC@NLO calculation has a sizable uncertainty from the choice of renormalization scale between 0.5 and 2 (+40\%,-25\%), from CTEQ PDF variations (+10\%,-6\%), and from the choice of b-quark mass between 4.5 and 5.0 GeV (+17\%,-14\%). The dominant scale uncertainty is overlaid as an uncertainty band in Fig.~\ref{fig:bjetdiffxsec}.

\section{Conclusions\label{sec:Conclusions}}

First measurements of $b$-quark inclusive cross sections were performed with the CMS experiments in p-p collisions at $\sqrt{s}=7$~TeV. The measurements utilize different techniques and cover a wide range of $b$-quark transverse momenta. The results were compared to NLO QCD predictions and various Monte Carlo models. The cross section measured with semileptonic decays into muons is above NLO QCD predictions at low momenta and central pseudorapidities. The $b$-jet cross section measured with secondary vertex tagging for $p_T>18$~GeV is in reasonable agreement with NLO QCD but exhibits shape differences at large rapidities. It is foreseen to extend the measured kinematic range by analyzing the larger data samples collected in the 2010 LHC run. 

\section*{Acknowledgements}

We thank M.Cacciari and P.Nason for the useful discussions and for providing the FONLL and POWHEG predictions.





\bibliographystyle{elsarticle-num}
\bibliography{<your-bib-database>}

\begin{thebibliography}{00}


%
%
\bibitem{Abachi:1994kj}
  S.Abachi {\it et al.}  [D0 Collaboration],
  Phys. Rev. Lett. {\bf 74}, 3548 (1995).
\bibitem{Abbott:1999se}
  B.~Abbott {\it et al.}  [D0 Collaboration],
  Phys. Lett. {\bf B487}, 264 (2000).
\bibitem{Abe:1993sj}
  F.~Abe {\it et al.}  [CDF Collaboration],
  Phys. Rev. Lett. {\bf 71}, 500 (1993).
\bibitem{Aaltonen:2009xn}
  T. Aaltonen {\it et al.}  [CDF Collaboration],
  Phys. Rev. {\bf D79}, 092003 (2009).
%
%
\bibitem{Adloff:1999nr}
 C.~Adloff {\it et al.}  [H1 Collaboration],
 Phys. Lett. {\bf B467}, 156 (1999).
\bibitem{Aktas:2005zc}
 A.~Aktas {\it et al.}  [H1 Collaboration],
 Eur. Phys. J. {\bf C41} 453 (2005).
\bibitem{Chekanov:2008tx}
 S.~Chekanov {\it et al.}  [ZEUS Collaboration],
 JHEP {\bf 04} 133 (2009).
\bibitem{Chekanov:2009kj}
 S.~Chekanov {\it et al.}  [ZEUS Collaboration],
 Eur. Phys. J. {\bf C65} 65 (2010).

%
%
\bibitem{Acciarri:2000kd}
M.~Acciarri {\it et al.}  [L3 Collaboration],
Phys. Lett {\bf B503} 10 (2001).
%
%
\bibitem{:2008zzk}
  R.~Adolphi {\it et al.}  [CMS Collaboration],
  JINST {\bf 3}, S08004 (2008).
\bibitem{BPH-10-007} CMS Collaboration, CMS-PAS-BPH-10-007 (2010).
\bibitem{Sjostrand:2006za}
  T.~Sjostrand, S.~Mrenna and P.~Z.~Skands,
  JHEP {\bf 0605}, 026 (2006)
  [arXiv:hep-ph/0603175].
\bibitem{Corcella:2002jc}
  G.~Corcella {\it et al.},
  arXiv:hep-ph/0210213.
\bibitem{Frixione:2003ei}
  S.~Frixione, P.~Nason and B.~R.~Webber,
  JHEP {\bf 0308}, 007 (2003)
  [arXiv:hep-ph/0305252].
\bibitem{Cacciari:2003uh}
  M.~Cacciari, S.~Frixione, M.~L.~Mangano, P.~Nason and G.~Ridolfi,
  JHEP {\bf 0407}, 033 (2004)
  [arXiv:hep-ph/0312132].
\bibitem{Frixione:2007vw}
  S.~Frixione, P.~Nason and C.~Oleari,
  JHEP {\bf 0711}, 070 (2007)
  [arXiv:0709.2092 [hep-ph]].
\bibitem{Pumplin:2002vw}
  J.~Pumplin, D.~R.~Stump, J.~Huston, H.~L.~Lai, P.~M.~Nadolsky and W.~K.~Tung,
  JHEP {\bf 0207}, 012 (2002)
  [arXiv:hep-ph/0201195].
%
%
\bibitem{BPH-10-009} CMS Collaboration, CMS-PAS-BTV-10-009 (2010).
\bibitem{Cacciari:2008gp}
  M.~Cacciari, G.~P.~Salam and G.~Soyez,
  JHEP {\bf 0804}, 063 (2008)
  [arXiv:0802.1189 [hep-ph]].
\bibitem{PFT-09-001} CMS Collaboration, CMS-PAS-PFT-09-001 (2009).
\bibitem{PFT-10-001} CMS Collaboration, CMS-PAS-PFT-10-001 (2010).
\bibitem{BTV-10-001} CMS Collaboration, CMS-PAS-BTV-10-001 (2010).
\bibitem{QCD-10-011} CMS Collaboration, CMS-PAS-QCD-10-011 (2010).
\end{thebibliography}



\end{document}